\documentstyle[12pt,fleqn]{article}
\def\con{{}_{\_\rule{-1pt}{0pt}\_}\rule{-2pt}{0pt}
\raise1.5pt\hbox{$\mid$}\hspace{2pt}}
\newtheorem{TH}{Theorem}
\newtheorem{LM}{Lemma}
\date{}
\title{\bf A Gauge-invariant Hamiltonian Description \\
of the Motion of Charged Test Particles}
\author{Dariusz Chru\'sci\'nski\footnotemark\\
        Institute of Physics,
        Nicholas Copernicus University\\
        ul. Grudzi\c{a}dzka 5/7, 87-100 Toru\'n, Poland\\
       and\\
       Jerzy Kijowski\footnotemark\\
       Centrum Fizyki Teoretycznej PAN\\
        Aleja Lotnik\'ow 32/46, 02-668 Warsaw, Poland}
\begin{document}
\maketitle

\begin{abstract}
New, gauge-independent, second-order Lagrangian for the motion of
classical, charged test particles is used to derive the corresponding 
Hamiltonian formulation. For this purpose a Hamiltonian description
of theories derived from the second-order Lagrangian is presented.
Unlike in the standard approach, the canonical momenta arising here are
explicitely gauge-invariant and have a clear physical interpretation.
The reduced symplectic form obtained this way is equivalent to Souriau's
form. This approach illustrates a new method of deriving equations of motion
from field equation.
\end{abstract}

\begin{sloppypar}

\def\thefootnote{\relax}\footnotetext{$^*$e-mail: darch@phys.uni.torun.pl}
\def\thefootnote{\relax}\footnotetext{$^\dagger$e-mail: 
kijowski@cft.edu.pl}
\section{Introduction}

In \cite{pracka1} a new method of deriving equations of motion from
field equations was proposed. The method is based on an analysis of the
geometric structure of generators of the Poincar\'e group and may by
applied to
any special-relativistic, lagrangian field theory.  In the case of
classical electrodynamics, this method leads uniquely to a manifestly
gauge-invariant, second order Lagrangian ${\cal L}$ for the motion of
charged test particles:
\begin{equation}
{\cal L} = L_{particle} + {\cal L}_{int} = - \sqrt{1-{\bf v}^2} \
( m  -  a^{\mu} u^{\nu} M_{\mu\nu}^{int}(t,{\bf q},{\bf v}) )
\  ,  \label{calL}
\end{equation}
where $u^{\mu}$ denotes the (normalized) four-velocity vector
\begin{eqnarray}
(u^\mu) = (u^0 , u^k) :=
\frac 1{\sqrt {1-{\bf v}^2}} (1,v^k)  \ ,
\end{eqnarray}
and $a^\mu := u^\nu \nabla_\nu u^\mu$ is the particle's acceleration
(we use the Heaviside-Lorentz system of units with the velocity of
light $c=1$). The skew-symmetric tensor $M_{\mu\nu}^{int}(t,{\bf
q},{\bf v})$ is equal to the amount of the angular-momentum of the
field, which is acquired by our physical system, when the Coulomb field
accompanying the particle moving with velocity ${\bf v}$ through the
space-time point $(t,{\bf q})$, is added to the background (external)
field. More precisely: the total energy-momentum tensor corresponding
to the sum of the background field $f_{\mu\nu}$ and the above Coulomb
field decomposes in a natural way into a sum of 1) terms quadratic in
the background field, 2) terms quadratic in the Coulomb field 3) mixed
terms. The quantity $M_{\mu\nu}^{int}$ is equal to this part of the
total angular-momentum $M_{\mu\nu}$, which we obtain integrating only
the mixed terms of the energy-momentum tensor.

The above result is a by-product of a consistent theory of interacting
particles and fields (cf. \cite{KIJ}, \cite{Kij-Dar}), called {\it
Electrodynamics of Moving Particles}.

We have proved in \cite{pracka1} that the new Lagrangian (\ref{calL})
differs from the standard one
\begin{equation}   \label{L}
L = L_{particle} + L_{int} = - \sqrt{1-{\bf v}^2} \
( m  - e u^{\mu}A_{\mu}(t,{\bf q}) )\ ,
\end{equation}
by (gauge-dependent) boundary corrections only. Therefore, both
Lagrangians generate the same equations of motion for test particles in
an external field. In the present paper we explicitly derive these
equations and construct the gauge-invariant Hamiltonian description of
this theory.

Standard Hamiltonian formalism, based on the gauge-dependent Lagrangian
(\ref{L}), leads to the gauge-dependent Hamiltonian
\begin{eqnarray}          \label{H}
H(t,{\bf q},{\bf p}) = \sqrt{m^2 + ({\bf p} + e{\bf A}(t,{\bf q}))^2} +
eA_0(t,{\bf q}) \ ,
\end{eqnarray}
where the gauge-dependent quantity
\begin{equation}   \label{p-standard}
p_k := p^{kin}_k - eA_k(t,{\bf q}) =  m u_k - eA_k(t,{\bf q})\
\end{equation}
plays role of the momentum canonically conjugate to the particle's
position $q^k$.

As was observed by Souriau (see \cite{Souriau}), we may replace the above
non-physical momentum in the description of the phase space of
this theory by the gauge-invariant quantity $p^{kin}$. The price we pay
for this change is, that the canonical contact form, corresponding to
the theory of free particles:
\begin{equation}
\Omega = dp^{kin}_\mu \wedge dq^\mu\ , \label{omega}
\end{equation}
has to be replaced by its deformation:
\begin{equation}  \label{omegaSou}
\Omega_S := \Omega - {e}\,f_{\mu\nu}\, dq^\mu \wedge dq^\nu\ ,
\end{equation}
where $e$ is the particle's charge.

Both $\Omega$ and $\Omega_S$ are defined on the ``mass-shell'' of the
kinetic momentum, i.~e.~on the surface $(p^{kin})^2 = - m^2$ in the
cotangent bundle $T^*M$ over the space-time $M$ (we use the Minkowskian
metric with the signature $(-,+,+,+)$). The forms contain the entire
information about dynamics: for free particles the admissible
trajectories are those, whose tangent vectors belong to the degeneracy
distribution of $\Omega$. Souriau noticed that replacing (\ref{omega})
by its deformation (\ref{omegaSou}) we obtain the theory of motion of
the particle in a given electromagnetic field $f_{\mu\nu}$.

The new approach, proposed in the present paper is based on Lagrangian
(\ref{calL}). It leads {\em directly} to a perfectly gauge-invariant
Hamiltonian, having a clear physical interpretation as the sum of two
terms: 1) kinetic energy $m u_0$ and 2) ``interaction energy'' equal to
the ammount of field energy acquired by our physical system, when the
particle's Coulomb field is added to the background field.

When formulated in terms of contact geometry, our approach leads uniquely
to a new form $\Omega_N$:
\begin{equation}  \label{omegaNew}
\Omega_N := \Omega - {e}\,h_{\mu\nu}\, dq^\mu \wedge dq^\nu\ ,
\end{equation}
where
\begin{equation}  \label{h}
h_{\mu\nu} := 2 ( f_{\mu\nu} -
u_{[\mu}f_{\nu]\lambda}u^\lambda)
\end{equation}
(brackets denote antisymmetrization), i.~e. we prove the following
\begin{TH}
The one dimensional degeneracies of the form $\Omega_N$ restricted to the
particle's ``mass-shell'' correspond to the trajectories of a test particle
moving in external electromagnetic field.
\end{TH}
It is easy to see that both
$\Omega_S$ and $\Omega_N$, although different, have the same degeneracy
vectors, because $h$ and $f$ give the same value on the velocity vector
$u_\nu$:
\begin{equation}
u^\nu h_{\mu\nu} = u^\nu f_{\mu\nu}\ . \label{uh-uf}
\end{equation}
Hence, both define the same equations of motion. We stress, however,
that our $\Omega_N$ is {\em uniquely obtained} from the gauge-invariant
Lagrangian (\ref{calL}) {\em via} the Legendre transformation.

The paper is organized as follows. In section~\ref{CanStruc} we sketch
briefly the (relatively little known) Hamiltonian formulation of
theories arising from the second order Lagrangian.  In
section~\ref{equations} we prove explicitly that the
Euler-Lagrange equations derived from $\cal L$ are equivalent to the
Lorentz equations of motion.  Finally, Section~\ref{HAMILTONIAN}
contains the gauge-invariant Hamiltonian structure of the theory.

\section{Canonical formalism for a 2-nd order
Lagrangian theory}
\label{CanStruc}

Consider a theory described by the 2-nd order lagrangian $L = L(q^i
,{\dot{q}}^i ,{\ddot{q}}^i )$ (to simplify the notation we will skip
the index ``$i$'' corresponding to different degrees of freedom $q^i$;
extension of this approach to higher order Lagrangians is
straightforward).  Introducing auxiliary variables $v = \dot{q}$ we can
treat our theory as a 1-st order one with lagrangian constraints $\phi
:= \dot{q} - v = 0$ on the space of lagrangian variables
$(q,\dot{q},v,\dot{v})$.  Dynamics is generated by the following
relation:
\begin{equation}   \label{rel1}
d\, L(q,v,\dot{v}) =
\frac{d}{dt}\big(p\,dq + \pi \,dv\big) = \dot{p}\,dq + p\,d\dot{q} +
\dot{\pi}\,dv + \pi \, d{\dot v} \ .
\end{equation}
where $(p, \pi)$ are momenta canonically conjugate to $q$ and $v$
respectively.
Because $L$ is defined only on the constraint submanifold, its
derivative $dL$ is not uniquely defined and has to be understood as a
collection of {\em all the covectors} which are compatible with the
derivative of the function along constraints. This means that the left
hand side is defined up to $ \mu (\dot{q} - v)$, where
$\mu$  are Lagrange multipliers corresponding to constraints $\phi =
0$ . We conclude that $p = \mu$ is an arbitrary covector and
(\ref{rel1}) is equivalent to the system of dynamical equations:
\begin{eqnarray}
 \pi &=& \frac{\partial L}{\partial \dot{v}} \, ,
\nonumber \\
\dot{p} &=& \frac{\partial L}{\partial q}\, ,
\nonumber \\
\dot{\pi} &=&
\frac{\partial L}{\partial v} - p \, .
\end{eqnarray}
The last equation implies the definition of the canonical momentum $p$:
\begin{equation}   \label{p}
p = \frac{\partial L} {\partial v} - \dot{\pi} = \frac{\partial L}
{\partial v} - \frac{d}{dt}\left( \frac{\partial L}{\partial
\dot{v}} \right)\ .
\end{equation}
We conclude, that equation
\begin{equation}
\dot{p} = \frac{d}{dt}\left(\frac{\partial L}{\partial v}\right) -
\frac{d^{2}}{dt^{2}}\left(\frac{\partial L}{\partial \dot{v}}
\right) \, . \end{equation}
is equivalent, indeed, to the  Euler-Lagrange equation:
\begin{equation} \label{EL}
\frac{\delta L}{\delta q} := \frac{d^{2}}{dt^{2}}\left(\frac
{\partial L}{\partial \dot{v}}\right) - \frac{d}{dt}\left(\frac
{\partial L}{\partial v}\right) + \frac{\partial L}{\partial q} = 0\, .
\end{equation}
The hamiltonian description (see e.~g.~\cite{Marsden}) is obtained from
the Legendre transformation applied to (\ref{rel1}):
\begin{equation}      \label{rel2}
-dH = \dot{p}\,dq - \dot{q}\,dp +\dot{\pi}\,dv -\dot{v}\,d\pi\, ,
\end{equation}
where $H(q,p,v,\pi) = p\,v +\pi \,\dot{v} - L(q,v, \dot{v} )$. In this
formula we have to insert $\dot{v} = \dot{v} (q,v,\pi )$, calculated
from  equation
$\pi = \frac{\partial L}{\partial
\dot{v}}$. Let us observe that $H$ is linear with respect to the
momentum $p$. This is a characteristic feature of the 2-nd order
theory.

In generic situation, Euler-Lagrange equations (\ref{EL}) are of 4-th
order. The corresponding 4 hamiltonian equations describe, therefore,
the evolution of $q$ and its derivatives up to third order.  Due to
Hamiltonian equations implied by relation (\ref{rel2}), the information
about succesive derivatives of $q$ is carried by $(v,\pi ,p)$:

\begin{itemize}

\item v describes $\dot{q}$  \begin{equation}
\dot{q} = \frac{\partial H}{\partial p} \equiv v \end{equation}
hence, the constraint $\phi = 0$ is reproduced due to linearity of $H$
with respect to $p$,

\item $\pi$ contains information about $\ddot{q}$:
\begin{equation}
\dot{v} = \frac{\partial H}{\partial \pi}\, , \end{equation}

\item   $p$ contains information about $\stackrel{...}{q}$
\begin{equation}
\dot{\pi} = - \frac{\partial H}{\partial v} = \frac{\partial
L}{\partial v} - p \, ,\end{equation}

\item  the true dynamical equation equals
\begin{equation}
\dot{p} = - \frac{\partial H}{\partial q} = \frac{\partial L}
{\partial q} \, . \end{equation}
\end{itemize}


\section{Equations of motion from the variational principle }
\label{equations}

In this section we explicitly derive the particle's equations of motion from
the variational principle based on the gauge-invariant Lagrangian
(\ref{calL}).
 The Euler-Lagrange equations for a second order Lagrangian theory are
given by
\begin{equation}  \label{E-L}
\dot{p}_k = \frac{\partial {\cal L}}{\partial q^{k}}\ ,
\end{equation}
where, as we have seen in the previous section, the momentum $p_k$
canonically conjugate to the particle's position $q^k$ is defined as:
\begin{equation}  \label{E-L1}
p_{k} :=  \frac{\partial {\cal L}}{\partial v^{k}} -  \dot{\pi}_k
\end{equation}
and
\begin{equation}          \label{pik}
\pi_k := \frac{\partial {\cal L}}{\partial
\dot{v}^{k}} =
\frac{1}{\sqrt{1-{\bf v}^2}}\ u^\nu M^{int}_{k\nu} (t,{\bf q},{\bf v})
\ .
\end{equation}
Now,
\begin{equation} \label{uM}
u^\nu M^{int}_{k\nu} = u^0 M^{int}_{k0} + u^l M^{int}_{kl} =
- u^0 r^{int}_k + u^l \epsilon_{kl}^{\ \ m} s^{int}_m\ ,
\end{equation}
where $r^{int}_k$ and $s^{int}_m$ are the static momentum and the
angular momentum of the interaction tensor. They are defined as
follows: we consider the sum of the (given) background field
$f_{\mu\nu}$ and the boosted Coulomb field ${\bf
f}_{\mu\nu}^{(y,u)}$ accompanying the
particle moving with constant four-velocity $u$ and passing through
the space-time point $y=(t,{\bf q})$. Being bi-linear in fields, the
energy-momentum tensor $T^{total}$ of the total field
\begin{equation}
f_{\mu\nu}^{total} := f_{\mu\nu} + {\bf f}_{\mu\nu}^{(y,u)}
\end{equation}
may be decomposed into three terms: the energy-momentum tensor of the
background field $T^{field}$, the Coulomb energy-momentum tensor
$T^{particle}$, which is composed of terms quadratic in
${\bf f}_{\mu\nu}^{(y,u)}$ and the ``interaction tensor'' $T^{int}$,
containing mixed terms:
\begin{equation}
T^{total} = T^{field} + T^{particle} + T^{int} \label{decomposition} \ .
\end{equation}
Interaction quantities (labelled with ``int'') are those obtained by
integrating appropriate components of $T^{int}$. Because all the three
tensors are conserved outside of the sources (i.~e.~outside of two
trajectories: the actual trajectory of our particle and the straight
line  passing through the space-time point $y$ with four-velocity
$u$), the integration gives the same result when performed over {\em
any} asymptoticaly flat Cauchy 3-surface passing through $y$.

In particular, $r^{int}$ and $s^{int}$ may be written in terms of the
laboratory-frame components of the electric and magnetic
fields as follows:
\begin{eqnarray}   \label{rk}
r^{int}_k(t,{\bf q},{\bf v}) &=&  \int_{\Sigma} d^3x\ (x_k - q_k)({\bf D}{\bf
D}_0     +
{\bf B}{\bf B}_0 )\ ,\\    \label{sk}
s^{int}_m(t,{\bf q},{\bf v})
&=& \epsilon_{mij} \int_{\Sigma} d^3x\ (x^i - q^i)( {\bf D} \times
{\bf B}_0 + {\bf D}_0 \times {\bf B})^j\ ,
\end{eqnarray}
where ${\bf D}$ and ${\bf B}$ are components of the external field $f$,
whereas
${\bf D}_0$ and ${\bf B}_0$ are components of ${\bf f}^{(y,u)}$, i.e.:
\begin{eqnarray}
{\bf D}_0({\bf x};{\bf q},{\bf v}) &=&
\frac{e}{4\pi |{\bf x}-{\bf q}|^3} \frac{1-{\bf v}^2}{\left(1-{\bf v}^2 +
\left(\frac{{\bf v}({\bf x}-{\bf q})}{|{\bf x}-{\bf
q}|}\right)^2\right)^{3/2}}\
({\bf x}-{\bf q})\ ,\label{Dzero} \\
{\bf B}_0({\bf x};{\bf q},{\bf v}) &=& {\bf v} \times
{\bf D}_0({\bf x};{\bf q},{\bf v})\ . \label{Bzero}
\end{eqnarray}

It may be easily seen that
quantities $r^{int}_k$ and $s^{int}_m$ are not independent. They fulfill the
following condition:
\begin{equation}   \label{s-r}
s^{int}_k = - \epsilon_{kl}^{\ \ m}v^lr^{int}_m\ .
\end{equation}
To prove this relation let us observe that in the particle's rest-frame
(see the Appendix for the definition) the
angular momentum corresponding to $T^{int}$ vanishes (cf.
\cite{pracka1}). When translated to the language of laboratory frame,
this is precisely equivalent to the above relation.

Inserting (\ref{s-r}) into (\ref{uM})  we finally get
\begin{equation}     \label{final-pik}
\pi_k = - \left(\delta^l_k + \frac{v^lv_k}{1-{\bf v}^2}\right)r^{int}_l\ .
\end{equation}
The quantity $r^{int}_k$ depends upon time {\em via} the time
dependence of the external fields $({\bf D}(t,{\bf x}),{\bf B}(t,{\bf
x}))$, the particle's position {\bf q} and the particle's velocity {\bf
v}, contained in formulae (\ref{Dzero}) -- (\ref{Bzero}) for the
particle's Coulomb field.

Now, we are ready to compute $p_k$ from (\ref{E-L1}):
\begin{eqnarray}
p_k &=& \frac{mv_k}{\sqrt{1-{\bf v}^2}} +  \dot{v}^l\frac{\partial
\pi_l}{\partial v^k}  - \left( \frac{\partial\pi_k}{\partial t}
+ v^l \frac{\partial\pi_k}{\partial q^l}
+ \dot{v}^l \frac{\partial\pi_k}{\partial v^l}  \right)\nonumber\\
&=&  p^{kin}_k - \left( \frac{\partial\pi_k}{\partial t}
+  v^l \frac{\partial\pi_k}{\partial q^l} \right)
- \dot{v}^l\left(\frac{\partial\pi_k}{\partial v^l}
-  \frac{\partial\pi_l}{\partial v^k} \right)\ .
\end{eqnarray}
Observe, that the  momentum $p_k$ depends upon time, particle's position and
velocity but also on particle's acceleration. Hovewer,
using (\ref{rk})  one easily shows
 that due to the
\begin{LM}   \label{LM1}   $\rule{0ex}{3ex}$
\begin{equation}  \label{pi-kl}
\frac{\partial\pi_k}{\partial v^l} -
\frac{\partial\pi_l}{\partial v^k} = 0\ ,
\end{equation}
\end{LM}
the term proportional to $\dot{v}^l$ vanishes (see Appendix for the proof).
Moreover, one can prove the following
\begin{LM}                 \label{LM2}  $\rule{0ex}{3ex}$
    \begin{equation}   \label{pi-kt}
 \frac{\partial\pi_k}{\partial t}
+ v^l \frac{\partial\pi_k}{\partial q^l} = - p^{int}_k\ ,
\end{equation}
\end{LM}
where we denote
\begin{equation}
p_k^{int}(t,{\bf q},{\bf v}) = \int_{\Sigma}d^3x\ ({\bf D}\times{\bf B}_0
+ {\bf D}_0\times{\bf B})_k \ .
\end{equation}
For the proof see Appendix.
We see that $p^{int}_k$ is the spatial part of the ``interaction
momentum'':
\begin{equation}  \label{p-int-mu}
p^{int}_\mu(t,{\bf q},{\bf v}) = \int_{\Sigma} T^{int}_{\mu\nu}\,
d\Sigma^\nu\ ,
\end{equation}
where $\Sigma$ is any hypersurface intersecting the particle's
trajectory at the point $(t,{\bf q}(t))$. The above integral is well
defined (cf.
\cite{KIJ}) and it is invariant with respect to changes of $\Sigma$, provided
the intersection point with the trajectory does not change. It was shown in
\cite{pracka1} that $p^{int}_\mu$ is orthogonal to the particle's
four-velocity, i.e. $p^{int}_\mu u^\mu =0$.

Finally, the momentum canonically conjugate to the particle's position
equals:
\begin{equation}  \label{pk}
p_k = p^{kin}_k + p^{int}_k(t,{\bf q},{\bf v}) \ .
\end{equation}
It is a sum of two terms: kinetic momentum $p^{kin}_k$ and the amount
of momentum $p^{int}_k$ which is acquired by our system, when the
particle's Coulomb field is added to the background (external) field.
We stress, that contrary to the standard formulation based on
(\ref{L}), our canonical momentum (\ref{pk}) is gauge-invariant.

Now, Euler-Lagrange equations (\ref{E-L}) read
\begin{equation}
\frac{d p^{kin}_k}{dt} + \frac{d p^{int}_k}{dt} =
\frac{\partial {\cal L}}{\partial q^k}\ ,
\end{equation}
or in a more transparent way:
\begin{equation}
\frac{d}{dt}\left(\frac{mv_k}{\sqrt{1-{\bf v}^2}}\right) =
- \left( \frac{\partial p^{int}_k}{\partial t}
+ v^l \frac{\partial p^{int}_k}{\partial q^l}\right)
- \dot{v}^l \left(\frac{\partial p^{int}_k}{\partial v^l}
-  \frac{\partial \pi_l}{\partial q^k}\right)\ .
\end{equation}
Again,
using definitions of $\pi_l$ and $p^{int}_k$ one shows that due to the
following
\begin{LM}  \label{LM3}   $\rule{0ex}{3ex}$
\begin{equation}   \label{pk-pil}
\frac{\partial p^{int}_k}{\partial v^l} -
\frac{\partial\pi_l}{\partial q^k} = 0\ .
\end{equation}
\end{LM}
 the
term proportional to the particle's acceleration vanishes (for the proof see
Appendix). The last step in our derivation is to calculate
$ \frac{\partial p^{int}_k}{\partial t}
  + v^l \frac{\partial p^{int}_k}{\partial q^l}$.
In the Appendix we show that the following identities hold:
\begin{LM}   \label{LM4}    $\rule{0ex}{3ex}$
\begin{equation}   \label{p-int-dt}
 \frac{\partial p^{int}_k}{\partial t}
  + v^l \frac{\partial p^{int}_k}{\partial q^l} = -
e\, \sqrt{1-{\bf v}^2}\ u^\nu f_{k\nu}(t,{\bf q}) = -
 e(E_k(t,{\bf q}) +
\epsilon_{klm}v^lB^m(t,{\bf q}))    \ .
\end{equation}
\end{LM}
Therefore,  the term
$ \frac{\partial p^{int}_k}{\partial t}
   + v^l \frac{\partial p^{int}_k}{\partial q^l}$
gives exactly the Lorentz force acting on a test
particle.

This way we proved that the Euler-Lagrange equations (\ref{E-L}) for the
variational problem based on $\cal L$ are equivalent
to the Lorentz equations for the motion  of charged particles:
\begin{equation}  \label{Lorentz}
\frac{d}{dt}\left(\frac{mv_k}{\sqrt{1-{\bf v}^2}}\right) =
 e(E_k(t,{\bf q}) +
\epsilon_{klm}v^lB^m(t,{\bf q}))    \ .
\end{equation}

\section{Hamiltonian formulation}
\label{HAMILTONIAN}

By Hamiltonian formulation of the theory we understand, usually, the
phase space of Hamiltonian variables ${\cal P}=(q,p)$ endowed with the
symplectic 2-form $\omega = dp\wedge dq$ and the Hamilton function $H$
(Hamiltonian) defined on $\cal P$. This function is interpreted as an
energy of the system. However, for time-dependent systems this
framework is usually replaced by (a slightly more natural) formulation
in terms of a {\em contact form}. For this purpose one considers the
{\it evolution space} ${\cal P} \times {\bf R}$ endowed with the {\it
contact} 2-form (i.e. closed 2-form of maximal rank):
\begin{equation}
\omega_H := dp\wedge dq - dH\wedge dt\ .
\end{equation}
In analytical mechanics this form, or rather its ``potential'' $pdq -
Hdt$, is called the {\it Poincar\'e-Cartan invariant}.  Obviously,
$\omega_H$ is degenerate on ${\cal P} \times {\bf R}$ and the
one-dimensional characteristic bundle of $\omega_H$ consists of the
integral curves of the system in ${\cal P} \times {\bf R}$. This kind
of description may be called the ``Heisenberg picture'' of classical
mechanics.  In this picture states are not points in $\cal P$ but
``particle's histories'' in ${\cal P}\times {\bf R}$ (see
\cite{Souriau}).

Let us construct the Hamiltonian structure for the theory based on our
second order Lagrangian $\cal L$. Let $\cal P$ denote the space of
Hamiltonian variables, i.e.  $({\bf q},{\bf p},{\bf v},\mbox{\boldmath
$\pi$})$, where {\bf p} and $\mbox{\boldmath $\pi$}$ stand for the
momenta canonically conjugate to {\bf q} and {\bf v} respectively.
Since our system is manifestly time-dependent (via the time dependence
of the external field) we pass to the evolution space endowed with the
contact 2-form
\begin{equation}    \label{Omega-calH}
\Omega_{{\cal H}} := dp_k \wedge dq^k + d\pi_k \wedge dv^k - d{\cal H}\wedge
dt\ ,
\end{equation}
where $\cal H$ denotes the time-dependent particle's Hamiltonian.

To find  $\cal H$ on ${\cal P} \times {\bf R}$ one
has to perform the (time-dependent) Legendre
transformation  $({\bf q},\dot{{\bf q}},{\bf v},\dot{{\bf v}}) \rightarrow
({\bf q},{\bf p},{\bf v},\mbox{\boldmath $\pi$})$, i.e. one has to calculate
$\dot{{\bf q}}$ and $\dot{{\bf v}}$ in terms of Hamiltonian variables from
formulae:
\begin{equation}
p_k =  \frac{\partial {\cal L}}{\partial \dot{q}^{k}}  - \dot{\pi}_k\ , \ \ \
\ \ \pi_k = \frac{\partial {\cal L}}{\partial \dot{v}^{k}}\ .
\end{equation}
This transformation is singular due to linear dependence of $\cal L$ on
$\dot{{\bf v}}$ and gives rise to the time-dependent constraints, given
by equations (\ref{pik}) and (\ref{pk}).  The constraints can be easily
solved i.e. momenta $p_k$ and $\pi_k$ can be uniquely parameterized by the
particle's position $q^k$, velocity $v^k$ and the time $t$.  Let ${\cal
P}^*$ denote the constrained submanifold of the evolution space ${\cal
P} \times {\bf R}$ parametrized by $({\bf q},{\bf p}^{kin},t)$. The reduced
Hamiltonian on ${\cal P}^*$ reads:
\begin{equation}
{\cal H}(t,{\bf q},{\bf v}) = p_kv^k + \pi_k\dot{v}^k - {\cal L} =
\frac{m}{\sqrt{1-{\bf v}^2}} + v^kp^{int}_k(t,{\bf q},{\bf v}).
\end{equation}
Due to identity $u^\mu p^{int}_\mu=0$ (cf. \cite{pracka1}) we have
\begin{equation}
{\cal H}(t,{\bf q},{\bf v}) = \frac{m}{\sqrt{1-{\bf v}^2}}
- p_0^{int}(t,{\bf q},{\bf v})\ ,
\end{equation}
and, therefore,
\begin{TH}
 The particle's Hamiltonian equals to the ``$-p_0$''
component of the following, perfectly gauge-invariant, four-vector
\begin{equation}
p_\mu := p^{kin}_{\mu} + p^{int}_\mu(t,{\bf q},{\bf v}) =
mu_\mu + p^{int}_\mu(t,{\bf q},{\bf v}) .
\end{equation}
\end{TH}
Using the laboratory-frame components of the
external electromagnetic field we get:
\begin{equation}      \label{p-int-0}
p^{int}_0(t,{\bf q},{\bf v}) =
- \int d^3x\ ({\bf D}{\bf D}_0 + {\bf B}{\bf B}_0)\ .
\end{equation}

Now, let us reduce the contact 2-form (\ref{Omega-calH}) on ${\cal P}^*$.
Calculating $p_k=p_k({\bf q},{\bf p}^{kin},t)$ and
$\pi_k=\pi_k({\bf q},{\bf p}^{kin},t)$ from (\ref{pik}) -- (\ref{pk}) and
inserting them
into (\ref{Omega-calH}) one obtains after a simple algebra:
\begin{equation}  \label{Omega1}
\Omega_N  =
dp^{kin}_\mu \wedge dq^\mu
- e \, h_{\mu\nu}\, dq^\mu \wedge dq^\nu\ ,
\end{equation}
where $q^0 \equiv t$ and
  $h_{\mu\nu}$ is  the following 4-dimensional tensor:
\begin{equation}   \label{G}
e \, h_{\mu\nu}(t,{\bf q},{\bf v}) :=
\frac{\partial p^{int}_\nu}{\partial q^\mu}
- \frac{\partial p^{int}_\mu}{\partial q^\nu} \ .
\end{equation}
Using techniques presented in the Appendix one easily proves
\begin{LM}      $\rule{0ex}{3ex}$
\begin{equation}
\frac{\partial p^{int}_\nu}{\partial q^\mu} =
e\Pi_\mu^{\ \lambda}f_{\lambda\nu}\ ,
\end{equation}
\end{LM}
where
\begin{equation}
\Pi_\mu^{\ \lambda} := \delta_\mu^{\ \lambda} + u_\mu u^\lambda
\end{equation}
is the projection on the hyperplane orthogonal to $u^\mu$ (i.e. to the
particle's rest-frame hyperplane, see the Appendix).  Therefore
\begin{equation}
h_{\mu\nu} =  \Pi_\mu^{\ \lambda}f_{\lambda\nu} -
\Pi_\nu^{\ \lambda}f_{\lambda\mu} = 2(f_{\mu\nu} -
u_{[\mu}f_{\nu]\lambda}u^\lambda)\ .
\end{equation}
where $a_{[\alpha}b_{\beta]} := \frac{1}{2}(a_\alpha b_\beta - a_\beta
b_\alpha)$. The form $\Omega_N$ is defined on a submanifold of
cotangent bundle $T^*M$ defined by the particle's ``mass shell'' $(p^{kin})^2
= - m^2$.

Observe, that the 2-form (\ref{Omega1}) has the same structure as the
Souriau's 2-form (\ref{omegaSou}).
They differ by the ``curvature'' 2-forms $f$ and $h$ only. However, the
difference ``$h - f$'' vanishes identically along the particle's trajectories
due to the fact that both
$f_{\mu\nu}$ and $h_{\mu\nu}$ have the same projections in the
direction of $u^\mu$ (see formula (\ref{uh-uf})).
We conclude that
the characteristic bundle of $\Omega_N$  and $\Omega_S$ are the
same and they are described by the following equations:
\begin{eqnarray}
\dot{q}^k &=& v^k\ ,\\
\dot{v}^k &=& \sqrt{1-{\bf v}^2}\ \frac{e}{m} (g^{kl} - v^kv^l)({E}_l +
\epsilon_{lij}v^i{B}^j)\ ,
\end{eqnarray}
which are equivalent to the Lorentz equations (\ref{Lorentz}).

We have two different contact structures which have the same
characteristic bundles. Therefore, from the physical point of view, these
forms are completely equivalent.



\def\theequation{A.\arabic{equation}}
\section*{Appendix}
\setcounter{equation}{0}

Due to the complicated dependence of the Coulomb field ${\bf D}_0$ and
${\bf B}_0$ on the particle's position {\bf q} and velocity {\bf v},
formulae containing the respective derivatives of these fields are
rather complex. To simplify the proofs, we shall use for calculations
the particle's rest-frame, instead of the laboratory frame.  The frame
associated with a particle moving along a trajectory $\zeta$ may be
defined as follows (cf. \cite{Kij-Dar},
\cite{pracka1}): at each point $(t,{\bf q}(t))\in \zeta$
we take the 3-dimensional hyperplane $\Sigma_t$ orthogonal to the
four-velocity $u^\mu$ (the {\it rest-frame hypersurface}).  We
parametrize $\Sigma_t$ by cartesian coordinates $(x^k),\ k=1,2,3$,
centered at the particle's position (i.e. the point $x^k=0$ belongs
always to $\zeta$). Obviously, there are infinitely many  such
coordinate systems on $\Sigma_t$, which differ from each other by an
$O(3)$-rotation. To fix uniquely coordinates $(x^k)$, we choose the
unique boost transformation relating the laboratory time axis
$\partial/\partial y^0$ with the four-velocity vector
$U:=u^\mu\frac{\partial}{\partial y^\mu}$.  Next, we define the position
of the $\partial/\partial x^k$ -- axis on $\Sigma_t$ by transforming
the corresponding $\partial/\partial y^k$ -- axis of the laboratory
frame by the same boost. The final formula relating Minkowskian
coordinates $(y^\mu)$ with the new parameters $(t,x^k)$ may be easily
calculated (see e.~g.~\cite{Kij-Dar}) from the above definition:
\begin{eqnarray}
y^0(t,x^l) & := &
 t +  \frac {1}{\sqrt{1 - {\bf v}^2(t)}} \ x^l v_l(t)
\ ,\nonumber
\\ y^k(t,x^l) & := & q^k(t) +
\left( {\delta}^k_l + \varphi ({\bf v}^2)v^k v_l \right) x^l
 \ , \label{embedding}
\end{eqnarray}
where we denote $\varphi (z):= \frac 1{z}
\left( \frac {1}{\sqrt{1 - {z}}} \ - 1 \right) =
\frac 1{\sqrt{1 -z} (1 + \sqrt{1-z} ) } $.

Observe, that the particle's Coulomb field
has in this co-moving frame extremely simple form:
\begin{equation}
\mbox{\boldmath ${\cal D}$}_0({\bf x}) =
\frac{e {\bf x}}{4\pi r^3}\ ,\ \ \ \
\mbox{\boldmath ${\cal B}$}_0({\bf x}) = 0\ ,
\end{equation}
where $r := |{\bf x}|$. That is why the calculations in this frame are
much easier than in the laboratory one.

Let ${\cal D}_k$ and ${\cal B}_k$ denote the rest-frame components of the
electric and magnetic field. They are related to $D_k$ and $B_k$ as follows:
\begin{eqnarray}  \label{D}
{\cal D}_k({\bf x},t;{\bf q},{\bf v}) &=&
\frac{1}{\sqrt{1-{\bf v}^2}} \left[ \left(\delta^l_k - \sqrt{1-{\bf v}^2}
\varphi({\bf v}^2)v^lv_k\right)D_l( y) -
\epsilon_{kij}v^iB^j(y) \right] ,\\
\label{B}
{\cal B}_k({\bf x},t;{\bf q},{\bf v}) &=&
\frac{1}{\sqrt{1-{\bf v}^2}} \left[ \left(\delta^l_k - \sqrt{1-{\bf v}^2}
\varphi({\bf v}^2)v^lv_k\right)B_l( y) +
\epsilon_{kij}v^iD^j(y) \right] ,
\end{eqnarray}
(the matrix $(\delta^l_k - \sqrt{1-{\bf v}^2}\,\varphi({\bf
v}^2)v^lv_k)$ comes from the boost transformation).

The field evolution with respect to the above {\it non inertial} frame is a
superposition of the following three transformations (cf.
\cite{pracka1}, \cite{KIJ},
\cite{Kij-Dar}):
\begin{itemize}
\item  time-translation in the direction of $U$,
\item  boost in the direction of the particle's acceleration $a^k$,
\item  purely spatial $O(3)$-rotation around the vector $\omega_m$,
\end{itemize}
where
\begin{eqnarray}   \label{ak}
a^k &:=& \frac{1}{1-{\bf v}^2}\left(\delta^k_l + \varphi({\bf
v}^2)v^kv_l\right)\dot{v}^l\ , \\
   \label{omegam}
\omega_m &:=& \frac{1}{\sqrt{1-{\bf v}^2}}\ \varphi({\bf
v}^2)v^k\dot{v}^l\epsilon_{klm}\ .
\end{eqnarray}
Therefore, the Maxwell equations  read (cf. \cite{KIJ}, \cite{Kij-Dar}):
\begin{eqnarray}  \label{Max-D}
\dot{{\cal D}}^n &=& \sqrt{1-{\bf v}^2}\ \frac{\partial}{\partial x^m}
\left[\left(\epsilon^{mk}_{\ \ \ i}{\cal D}^n
- \epsilon^{nk}_{\ \ i}{\cal D}^m\right)\omega_kx^i
- \epsilon^{mn}_{\ \ \ k}(1+a^ix_i){\cal B}^k\right]\ ,\\
\label{Max-B}
\dot{{\cal B}}^n &=& \sqrt{1-{\bf v}^2}\ \frac{\partial}{\partial x^m}
\left[\left(\epsilon^{mk}_{\ \ \ i}{\cal B}^n
- \epsilon^{nk}_{\ \ i}{\cal B}^m\right)\omega_kx^i
+ \epsilon^{mn}_{\ \ \ k}(1+a^ix_i){\cal D}^k\right]\ ,
\end{eqnarray}
(the factor $\sqrt{1-{\bf v}^2}$ is necessary, because the time $t$, which
we used to parametrize the particle's trajectory, is not a proper time along
$\zeta$ but the laboratory time).

On the other hand, the time derivative with respect to the co-moving frame
may be written as
\begin{equation}
\frac{d}{dt} = \frac{\partial}{\partial t} + v^k\frac{\partial}{\partial q^k}
+ \dot{v}^k\frac{\partial}{\partial v^k}
= \left(\frac{\partial}{\partial t}\right)_U
+ \dot{v}^k\frac{\partial}{\partial v^k}\ .
\end{equation}
Therefore, taking into account (\ref{Max-D}) and (\ref{Max-B}) we obtain:
\begin{eqnarray}   \label{D-t}
\left(  \frac{\partial}{\partial t} \right)_U {\cal D}^n
 &=&  \sqrt{1-{\bf v}^2}\ \epsilon^{nmk}\partial_m{\cal B}_k\ , \\
\label{B-t}
\left(  \frac{\partial}{\partial t} \right)_U {\cal B}^n
 &=& - \sqrt{1-{\bf v}^2}\ \epsilon^{nmk}\partial_m{\cal D}_k\ ,
\end{eqnarray}
and
\begin{eqnarray}  \label{D-v}
\frac{\partial}{\partial v^l}\ {\cal D}^n
&=& \sqrt{1-{\bf v}^2}\ \partial_m
\left[  \frac{\partial \omega_k}{\partial\dot{v}^l}
\left(  \epsilon^{mk}_{\ \ \ i}{\cal D}^n
- \epsilon^{nk}_{\ \ i}{\cal D}^m\right) x^i
-  \frac{\partial a^k}{\partial\dot{v}^l}
\epsilon^{mn}_{\ \ \ k}x_i{\cal B}^k\right]\ ,\\
\label{B-v}
\frac{\partial}{\partial v^l}\ {\cal B}^n
&=& \sqrt{1-{\bf v}^2}\ \partial_m
\left[  \frac{\partial \omega_k}{\partial\dot{v}^l}
\left(  \epsilon^{mk}_{\ \ \ i}{\cal D}^n
- \epsilon^{nk}_{\ \ i}{\cal D}^m\right) x^i
+  \frac{\partial a^k}{\partial\dot{v}^l}
\epsilon^{mn}_{\ \ \ k}x_i{\cal B}^k\right]\ .
\end{eqnarray}
To calculate the derivatives of ${\cal D}^k$ and ${\cal B}^k$ with respect to
the particle's position observe, that
\begin{equation}
\frac{\partial}{\partial y^k} = - \frac{v_k}{\sqrt{1-{\bf v}^2}}\ U
+ \left(\delta^i_k + \varphi({\bf
  v}^2)v^iv_k\right)\frac{\partial}{\partial x^i} \ .
\end{equation}
Therefore
\begin{eqnarray}     \label{D-q}
\frac{\partial}{\partial q^k}\ {\cal D}^n &=&
- \frac{v_k}{\sqrt{1-{\bf v}^2}}\ \epsilon^{nmi}\partial_m{\cal B}_i
+ \left(\delta^i_k + \varphi({\bf
  v}^2)v^iv_k\right)\partial_i{\cal D}^n \ , \\
\label{B-q}
\frac{\partial}{\partial q^k}\ {\cal B}^n &=&
 \frac{v_k}{\sqrt{1-{\bf v}^2}}\ \epsilon^{nmi}\partial_m{\cal D}_i
+ \left(\delta^i_k + \varphi({\bf
  v}^2)v^iv_k\right)\partial_i{\cal B}^n \ .
\end{eqnarray}
Now, using (\ref{D-t})--(\ref{B-v}) and (\ref{D-q})--(\ref{B-q}) we prove
Lemmas \ref{LM1}--\ref{LM4}.
\\

\noindent{\bf 1. Proof of Lemma~\ref{LM1}:}

Observe, that ``interaction static moment'' in the particle's rest frame
reads:
\begin{equation}
R^{int}_k := \int_{\Sigma_t}  x_k
(\mbox{\boldmath ${\cal D}$}_0 \mbox{\boldmath ${\cal D}$} +
\mbox{\boldmath ${\cal B}$}_0 \mbox{\boldmath ${\cal B}$})\,d^3x =
\frac{e}{4\pi}\int_{\Sigma_t} \frac{x_kx^i}{r^3}{\cal D}^i\, d^3x\ .
\end{equation}
Taking into account that
\begin{equation}
r^{int}_k = \frac{1}{\sqrt{1-{\bf v}^2}}
 \left(\delta^i_k - \sqrt{1-{\bf v}^2}\
\varphi({\bf v}^2)v^iv_k\right) R^{int}_i
\end{equation}
we obtain the formula for $\pi_k$ in terms of $R^{int}_i$:
\begin{equation}   \label{pi-R}
\pi_k = - \frac{1}{\sqrt{1-{\bf v}^2}} \left(\delta^i_k +
\varphi({\bf v}^2)v^iv_k\right)R^{int}_i\ .
\end{equation}
Now, using (\ref{D-v}) one gets:
\begin{equation}  \label{Rk-v}
\frac{\partial}{\partial v^l}\ R^{int}_i = \sqrt{1-{\bf v}^2}
\left\{ \frac{\partial a^m}{\partial\dot{v}^l}{\bf X}_{im} -
\frac{\partial\omega^m}{\partial\dot{v}^l}\epsilon_{im}^{\ \ \ j}R^{int}_j
\right\}\ ,
\end{equation}
where
\begin{equation}
\label{X-im}
{\bf X}_{im} =
\frac{e}{4\pi} \epsilon_{ijk}
\int_{\Sigma_t} \frac{x^jx_m}{r^3}{\cal B}^k\, d^3x\ .
\end{equation}
Therefore
\begin{equation}
\frac{\partial\pi_k}{\partial v^l} - \frac{\partial\pi_l}{\partial v^k}
= A^i_{\ kl}R^{int}_i - B^{im}_{kl}{\bf X}_{im}\ ,
\end{equation}
where
\begin{eqnarray}
A^i_{\ kl} &=&
\frac{\partial}{\partial v^k}\left[ \frac{1}{\sqrt{1-{\bf v}^2}}
 \left(\delta^i_l + \varphi({\bf v}^2)v^iv_l\right)\right]
- \frac{\partial}{\partial v^l}\left[ \frac{1}{\sqrt{1-{\bf v}^2}}
   \left(\delta^i_k + \varphi({\bf v}^2)v^iv_k\right)\right]
\nonumber\\  &+& \epsilon^i_{\ jm}\left[
\left(\delta^j_k + \varphi({\bf v}^2)v^jv_k\right)\frac{\partial\omega^m}
{\partial\dot{v}^l}  -
\left(\delta^j_l + \varphi({\bf v}^2)v^jv_l\right)\frac{\partial\omega^m}
{\partial\dot{v}^k} \right]\ ,\\
\label{Bimkl}
B^{im}_{kl} &=&
\left(\delta^i_k + \varphi({\bf v}^2)v^iv_k\right)
\frac{\partial a^m}{\partial\dot{v}^l} -
\left(\delta^i_l + \varphi({\bf v}^2)v^iv_l\right)
\frac{\partial a^m}{\partial\dot{v}^k} \nonumber\\ &=&
(1-{\bf v}^2) \left(
\frac{\partial a^i}{\partial\dot{v}^k}
\frac{\partial a^m}{\partial\dot{v}^l} -
\frac{\partial a^i}{\partial\dot{v}^l}
\frac{\partial a^m}{\partial\dot{v}^k} \right)\ .
\end{eqnarray}
Using the following properties of the function $\varphi(z)$:
\begin{eqnarray}        \label{I}
2\varphi(z) - (1-z)^{-1} + z\varphi^2(z) &=& 0\ ,\\
\label{II}
2\varphi'(z) - (1-z)^{-1}\varphi(z) - \varphi^2(z) &=& 0\ ,
\end{eqnarray}
one easily shows that $A^i_{\ kl} \equiv 0$. Moreover, observe
that $B^{im}_{kl}$ defined in (\ref{Bimkl}) is antisymmetric in $(im)$.
Therefore, to prove (\ref{pi-kl}) it is sufficient to show that the quantity
${\bf X}_{im}$
 is symmetric in $(im)$.
Taking into account that ${\cal B}^k = \epsilon^{klm}\partial_l{\cal A}_m$,
where
${\cal A}_m$ stands for the rest-frame components of
vector potential, one immediatelly gets:
\begin{equation}
\epsilon_{ijk}
\int_{\Sigma_t} \frac{x^jx_m}{r^3}{\cal B}^k\, d^3x
= \int_{\Sigma_t} r^{-5}({\cal A}_kx^k)(3x_ix_m - r^2g_{im})\,d^3x\ ,
\end{equation}
which ends the proof of (\ref{pi-kl}).
\\

\noindent{\bf 2. Proof of Lemma~\ref{LM2}:}

To prove (\ref{pi-kt}) observe that
\begin{equation}
\frac{\partial \pi_k}{\partial t} + v^l\frac{\partial \pi_k}{\partial q^l}
= \left(\frac{\partial}{\partial t}\right)_U \pi_k
= -  \frac{1}{\sqrt{1-{\bf v}^2}} \left(\delta^i_k +
  \varphi({\bf v}^2)v^iv_k\right)
\left(\frac{\partial}{\partial t}\right)_U   R^{int}_i\ .
\end{equation}
Now, using (\ref{D-t}) we obtain
\begin{eqnarray}
\lefteqn{
\left(\frac{\partial}{\partial t}\right)_U   R^{int}_i
\ = \ \sqrt{1-{\bf v}^2}
\int_{\Sigma_t} \frac{x_ix_k}{r^3}\epsilon^{kjm}\partial_j{\cal B}_m\, d^3x\
} \nonumber\\
&=& \sqrt{1-{\bf v}^2}
\int_{\Sigma_t}
\partial_j\left(\frac{x_ix_k}{r^3}\epsilon^{kjm}{\cal B}_m\right) d^3x
+ \sqrt{1-{\bf v}^2}
\int_{\Sigma_t} \epsilon_i^{\ km}\frac{x_k}{r^3}{\cal B}_m\,d^3x\ .
\end{eqnarray}
Due to the Gauss theorem
\begin{equation}
\int_{\Sigma_t}
\partial_j\left(\frac{x_ix_k}{r^3}\epsilon^{kjm}{\cal B}_m\right) d^3x
= \int_{\partial\Sigma_t}\frac{x_jx_ix_k}{r^4}\epsilon^{kjm}{\cal
B}_m \,d\sigma \equiv 0\ ,
\end{equation}
where $d\sigma$ denotes the surface measure on $\partial\Sigma_t$.
Moreover, observe that ``interaction momentum'' in the particle's rest-frame
reads:
\begin{equation}
P^{int}_i := \epsilon_{ikm} \int_{\Sigma_t} (
{\cal D}_0^k{\cal B}^m + {\cal D}^k{\cal B}_0^m)\,d^3x =
\frac{e}{4\pi} \epsilon_{ikm} \int_{\Sigma_t}
 \frac{x^k}{r^3}{\cal B}^m\,d^3x\   .
\end{equation}
Therefore
\begin{equation}
\left(\frac{\partial}{\partial t}\right)_U   R^{int}_i
= \sqrt{1-{\bf v}^2}\ P^{int}_i\ .
\end{equation}
Using the relation between $p^{int}_k$ and $P^{int}_i$
\begin{equation}    \label{p-P}
p^{int}_k =
\left(\delta^i_k + \varphi({\bf v}^2)v^iv_k\right) P^{int}_i
\end{equation}
we finally get (\ref{pi-kt}).
\\

\noindent{\bf 3. Proof of Lemma~\ref{LM3}:}

Using (\ref{B-v}) and (\ref{D-q}) we obtain:
\begin{eqnarray}
\frac{\partial}{\partial v^l}\ P^{int}_i &=&  \sqrt{1-{\bf v}^2}
\left\{
 \frac{\partial\omega^m}{\partial \dot{v}^l}\ \epsilon_{im}^{\ \ \
j}P^{int}_j
- \frac{\partial a^m}{\partial \dot{v}^l} {\bf Y}_{mi}
\right\}\ ,\\
\frac{\partial}{\partial v^l}\ R^{int}_i &=&
\frac{v_l}{\sqrt{1-{\bf v}^2}}\ P^{int}_i
+ \left(\delta^j_l + \varphi({\bf v}^2)v^jv_l\right){\bf Y}_{ij}\ ,
\end{eqnarray}
where
\begin{equation}
{\bf Y}_{ij}:=
\frac{e}{4\pi}\int_{\Sigma_t} \frac{x_i}{r^5}(3x^kx_j - r^2\delta^k_j){\cal
D}_k\,d^3x\ .
\end{equation}
Now, taking into account (\ref{pi-R}) and (\ref{p-P}) we have
\begin{equation}
\frac{\partial p^{int}_k}{\partial v^l} - \frac{\partial\pi_l}{\partial q^k}
= C^i_{\ kl} P^{int}_i\ ,
\end{equation}
where
\begin{eqnarray}
C^i_{\ kl} &:=& \frac{\partial}{\partial v^l}
\left(\delta^i_k + \varphi({\bf v}^2)v^iv_k\right)
- \sqrt{1-{\bf v}^2} \left(\delta^j_k + \varphi({\bf v}^2)v^jv_k\right)
 \frac{\partial\omega^m}{\partial \dot{v}^l}\ \epsilon_{jm}^{\ \ \ i}
\nonumber\\   &+&
\frac{v_k}{1-{\bf v}^2}\left(\delta^i_l + \varphi({\bf v}^2)v^iv_l\right)\ .
\end{eqnarray}
One easily shows that
due to properties (\ref{I})--(\ref{II}) $C^i_{\ kl}\equiv 0$, which ends the
proof of (\ref{pk-pil}).
\\

\noindent{\bf 4. Proof of Lemma~\ref{LM4}:}

Finally, to prove (\ref{p-int-dt}) let us observe that
\begin{equation}
\frac{\partial p^{int}_k}{\partial t} +
v^l\frac{\partial p^{int}_k}{\partial q^l} =
\left( \frac{\partial}{\partial t}\right)_U p^{int}_k =
\left(\delta^i_k + \varphi({\bf v}^2)v^iv_k\right)
\left(\frac{\partial}{\partial t}\right)_U P^{int}_i \ .
\end{equation}
Now, due to (\ref{B-t}) we get
\begin{eqnarray}
\lefteqn{
\left(  \frac{\partial}{\partial t} \right)_U P^{int}_i
  =  \sqrt{1-{\bf v}^2}\
\frac{e}{4\pi}\int_{\partial\Sigma_t} \frac{1}{r^2}{\cal D}_i\,d\sigma }
\nonumber\\   &=&
 - \sqrt{1-{\bf v}^2}\ \frac{e}{4\pi}\ \lim_{r_0\rightarrow 0} \int_{S(r_0)}
\frac{1}{r^2}{\cal D}_i\,d\sigma = - \sqrt{1-{\bf v}^2}\ e{\cal D}_i(t,0)\ ,
\end{eqnarray}
where
we choose
as two pieces of a boundary $\partial\Sigma_t$ a sphere at infinity and a
sphere
$S(r_0)$.
 Using the
fact that in
 the Heaviside-Lorentz system of units ${\cal D}_k = {\cal E}_k$ and
taking into account the formula (\ref{D})
we finally obtain   (\ref{p-int-dt}).

\end{sloppypar}
\end{document}